\documentclass[12pt]{article}
\usepackage{amsmath, amsfonts,euscript,bbm}
\usepackage{epsfig, cite}
\usepackage{color}
\usepackage{ifthen}
\usepackage{graphicx}
\usepackage{setspace} 
\newboolean{Notes}\setboolean{Notes}{true}
\newcommand{\notes}[1]{\ifthenelse{\boolean{Notes}}{\textcolor{black}{#1}}{}}

\newcommand{\skyp}[1]{}

\begin{document}

\bigskip
\hskip 4in\vbox{\baselineskip12pt \hbox{FERMILAB-PUB-08-550-A-T}  }
\bigskip\bigskip\bigskip

\centerline{\Large Holographic Geometry and Noise in Matrix Theory}
\bigskip
\bigskip
\centerline{\bf Craig J. Hogan$^{1,2}$ and Mark G. Jackson$^{1,3,4}$}
\medskip
\centerline{$^1$Particle Astrophysics Center}
\centerline{Fermi National Accelerator Laboratory}
\centerline{Batavia, Illinois 60510}
\medskip
\centerline{$^2$Enrico Fermi Institute, Kavli Institute for Cosmological Physics, }
\centerline{and Department of Astronomy and Astrophysics,}
\centerline{University of Chicago}
\centerline{Chicago, Illinois 60637}
\medskip
\centerline{$^3$ Theory Group}
\centerline{Fermi National Accelerator Laboratory}
\medskip
\centerline{$^4$Lorentz Institute for Theoretical Physics}
\centerline{University of Leiden}
\centerline{Leiden 2333CA, The Netherlands}
\bigskip
\bigskip
\begin{abstract}
Using Matrix Theory  as a concrete example of a fundamental holographic theory, we show that the emergent macroscopic spacetime displays a new macroscopic quantum structure, holographic geometry, and a new observable phenomenon, holographic noise, with phenomenology similar to that previously derived on the basis of a quasi-monochromatic wave theory.
Traces of matrix operators on a light sheet with a compact dimension  of size $R$ are interpreted as transverse position operators  for macroscopic bodies.  
An effective quantum wave equation for spacetime is derived from the Matrix Hamiltonian.  Its solutions display eigenmodes that connect longitudinal separation and transverse position operators on macroscopic scales.   Measurements of transverse relative positions of macroscopically separated bodies, such as signals in Michelson interferometers, are shown to display holographic nonlocality, indeterminacy and noise, whose properties can be predicted with no parameters except $R$.  Similar results are derived using a detailed scattering calculation of the matrix wavefunction.  Current experimental technology will allow a definitive and precise test or validation of  this interpretation  of holographic fundamental theories.  In the latter case,  they will yield   a direct measurement of $R$ independent of the gravitational definition of the Planck length, and a direct measurement of the total number of degrees of freedom.

\end{abstract}

\newpage            
\baselineskip=18pt
\section{Introduction}
If the world as a whole is a  quantum system, there should  be a quantum description of observables that correspond to properties of spacetime.  Arguments from gravitational thermodynamics \cite{'tHooft:1993gx,Susskind:1994vu} and superstring theory \cite{Maldacena:1997re,Witten:1995ex} suggest that the quantum description of spacetime may be holographic, that is, expressible as a local quantum theory with only two large spatial dimensions and a finite minimum length.  The large dimensions are interpreted as null surfaces or wavefronts in a dual, emergent three-dimensional world. Such theories are strongly motivated because by construction they respect holographic scaling of entropy and information \cite{Bousso:1999xy, Bousso:2002ju}.  Holographic degrees of freedom statistically explain the laws of gravitational thermodynamics \cite{Bekenstein:1972tm,Bardeen:gs,Bekenstein:1973ur, bek2, Hawking:1975sw,Jacobson:1995ab}: for example, the null surface represented by a black hole event horizon encodes the states of the hole, so that the total entropy is given by one quarter of the area in Planck units.

 Recently   it  has been conjectured \cite{Hogan:2007pk,Hogan:2008zw} that theories of this type generically create a macroscopic holographic quantum geometry with a new,  particular and precisely defined form of nonlocal  correlation:  an indeterminacy in transverse position that grows with scale $L$ like $\sqrt{LR}$, where $R$ is some fundamental length scale.  Arguments based on wave mechanics, rather than a fundamental theory, were used to derive this generic phenomenology.  In this paper we show concretely how exactly such a macroscopic holographic quantum geometry arises in a specific candidate fundamental theory that includes quantum spacetime:
 Matrix Theory \cite{Banks:1996vh}, a realization of M-theory, the strongly-coupled version of string theory \cite{Witten:1995ex}.  We derive from Matrix Theory an effective wave theory, including an expression for  the fundamental   quantum eigenmodes of the spacetime wavefunction.   This macroscopic quantum structure is new; it shows explicitly how the quantum degrees of freedom of spacetime are holographically encoded, and how they manifest themselves in the three dimensional world.  They lead to new nonlocal correlations, including  a new and precisely characterised observable phenomenon, ``holographic noise''.  
 
In the following section, we  adopt an interpretation of Matrix Theory in terms of macroscopic position observables. We  use wave mechanics to derive the low-frequency, large-longitudinal-separation limit of the theory and identify the quantum eigenstates of macroscopic spacetime, which describe   the observable relative positions of macroscopic bodies.  These eigenfunctions show a new relation between longitudinal and transverse position observables, different from the classical limit derived from field theory.  Position wavefunctions are shown to display the same surprisingly large transverse indeterminacy that was previously derived in \cite{Hogan:2007pk,Hogan:2008zw} using a model based on classical wave correlations. This result supports the conjecture that holographic quantum geometry, characterised by large transverse indeterminacy, is a generic prediction of holographic unified quantum theories. In section III we support the assertions used to construct this wave model--- in particular, the interpretation of matrix traces as transverse position operators---  by evaluating the properties of a scattered wavefunction in Matrix theory. 

The measurable phenomenon of holographic noise can be used to either rule out this interpretation of the physical meaning of Matrix theory,  or validate it.  In the latter case it will be possible to pursue an experimental program that studies properties of quantum geometry directly, including precise measurement of the fundamental frequency or minimum time.  
\section{Macroscopic Wave Theory from Matrix Theory}
\subsection{Matrix Theory}

Although Matrix Theory fully describes the dynamics occurring in 10 spatial dimensions, the native formulation is on a nine-dimensional light sheet.  The tenth (or ``M'') dimension is a longitudinal one compactified with size $R$ and thereby containing an infrared cutoff.  This scale corresponds to the 11-dimensional Planck scale $M_{\rm pl} ^{-1}$ as $R M_{\rm pl} = g_s^{2/3}$, where $g_s$ is the string coupling constant, but in our holographic noise theory it will be found to define the inverse ultraviolet frequency cutoff in the time dimension--- the minimum unit of time and length in the emergent macroscopic spacetime.

Matrix Theory considers the dynamics of $N$ D0-branes \cite{Polchinski:1995mt}, each having one unit of Kaluza-Klein (KK) momentum in the M-direction of compactification radius $R$, which we will call $z$.  By making $N$ very large the total momentum in this direction will be much larger than the momentum in any other direction. The highly-boosted system can then naturally be described by a Lagrangian used to propagate along the lightlike direction $z^+ \equiv (t+ z)/2$ and which defines the normal direction to a light sheet in the emergent holographic space:
\begin{equation}
\label{lagrangian}
L = {\rm tr} \   \sum_i \left\{ \frac{ {\dot X}^i {\dot X}^i }{ 2R} - {RM_{\rm pl}^6 \over 4} \sum_j [X^i,X^j]^2 \right\} - R M_{\rm pl} V .
\end{equation}
Here the fields $X^i$ are $N\times N$ matrices representing the D0-branes' position in the transverse directions within the sheet.  We have included a potential term $V$ for future use, but have neglected all supersymmetry-completing fermions and gauge field couplings since these will not be relevant to our conclusion.  The position of a macroscopic body (composed of the D0-branes) in the transverse $x^i$-direction is then given by $x^i=X^i_{c.m.}={1 \over N} {\rm tr}\ X^i$.  

In the case of a single D0-brane separated from $N \rightarrow \infty$ others, we may use standard perturbation theory treat the $X^i$ as (matrix-valued) quantum fields and construct an effective Lagrangian to describe their long-distance interactions \cite{Douglas:1996yp}, 
\begin{equation}
\label{leff}
 L_{\rm eff} =  \left[ \frac{v^2}{2R} - \alpha N M_{\rm pl}^{-9} \frac{v^4}{R^3 r^7}  \right] 
 \end{equation}
where $v^i \equiv dX^i/dz^+$ and $\alpha$ is some constant of order unity, and $r$ is the interaction distance. This is the expected 9+1-dimensional velocity-dependent Coulombic-like potential, the leading term appearing at $\mathcal O(v^4)$ due to supersymmetry.  Without explicitly solving the Schr\"odinger equation to obtain a wavefunction, we can still use (\ref{leff}) to glean the important scaling properties by observing how each quantity must scale to obtain a dimensionless action:
\begin{equation}
\label{scaling}
 X^i \sim  \frac{N^{1/9}} {M_{\rm pl}} , \hspace{0.5in} z^+ \sim \frac{N^{2/9}}{RM_{\rm pl}^2} , \hspace{0.5in} v \sim \frac{X}{z^+} \sim N^{-1/9} RM_{\rm pl} . 
 \end{equation}
This incompressibility demonstrates the `one quantum per transverse Planck volume' property expected of holography.  It is important because while there is no explicit UV cutoff in the sheet, there is an effective coupling among the internal brane degrees of freedom that limits the number of actual independent degrees of freedom in the emergent space to about one per Planck area.

\subsection{Effective Wave Theory for Spacetime Position Wavefunction}

We take two approaches in this paper to construct a macroscopic quantum limit of  Matrix theory. The simplest approach, which also makes the clearest connection to phenomenology, formulates a conjecture in the form of an effective theory: a wave equation relating longitudinal and transverse position.  Traces of matrices are posited, rather then demonstrated, to behave like Schr\"odinger wave operators in the macroscopic system. In the second approach (section III below) we use the full Matrix formulation to scatter a wave packet into a different direction. It confirms in general terms the behavior of this effective theory.

Holographic geometry arises in the following interpretation of Matrix theory in terms of conventional position variables in the emergent macroscopic Minkowski space.  
We first rewrite the Hamiltonian corresponding to the Lagrangian (\ref{lagrangian}) with the constant $\hbar$ restored, and without any scattering potential:
\begin{equation}\label{hamilton}
H= {R\over \hbar}\  {\rm tr}\ \sum_i \left\{{\Pi_i\Pi_i\over 2} + M_{\rm pl}^6 \sum_j {1\over 4} [X^i,X^j]^2 \right\}.
\end{equation}
Planck's constant is restored here to point out later that  it does not appear in the experimental predictions, and that the fluctuations are not the same as zero point quantum fluctuations in field theory.
We have defined the (matrix-valued) transverse momentum operator ${\Pi}_i  \rightarrow -i \hbar \frac{\partial}{\partial X^i}$ (note, however, that $H$ is simply a scalar).  Again,  $R$ denotes the radius of the  compactified dimension $z$, a fundamental length scale. This  direction becomes the longitudinal or normal direction to a light sheet  in the emergent macroscopic  holographic space.   Of the nine transverse directions in the full theory, we will be concerned only with the two large dimensions that survive at macroscopic scales. In a Michelson interferometer, we measure only one of these; the ``second'' arm is identified with the longitudinal dimension.  The macroscopic   behavior, describing transverse displacements at large longitudinal separation,  is dominated by the first (kinetic) term in Eq. (\ref{hamilton}).  Gradients in the wavefront    encode  position in the longitudinal direction, similar to the way parallax is encoded in an interferometric laboratory hologram.  We consider quasi-classical systems where  the effect of other terms, that encode the rest of  physics, including the familiar quantum mechanics of bodies and particles, can be neglected. That is, an effective theory of position can be separated from the local quantum mechanics, to isolate and study just the nonlocal, macroscopic properties of the holographic quantum geometry itself.  In this effective theory the two macroscopic $X^i$  are not coupled with each other. In what follows we thus suppress the indices on $X^i$ since the wavefunction in each transverse direction is independent.  For the Michelson system we  consider  only two dimensions, the longitudinal coordinate $z^+$ (from which emerges the virtual spatial $z$ coordinate normal to the light sheet in the lab frame), and one transverse direction $x$. 

In this limit the  Hamiltonian converts into wave language described by familiar Schr\"odinger wave operators.  The trace  $x=X_{c.m.}= \frac{1}{N} {\rm tr}\ X$ averages over the microscopic degrees of freedom so that the position is no longer a matrix, but becomes an ordinary position operator $x$.  
The conjugate momentum ${\rm tr} \Pi$  can be represented in the usual way by the operator ${\rm tr}\Pi=p= -i \hbar \partial/\partial x$. The Hamiltonian operator is represented in the usual way by
$H=i\hbar \partial/\partial z^+$.  
The evolution equation for the wavefunction $\psi(z^+,x)$ then becomes 
\begin{equation}\label{evolution}
\left( i \frac{\partial}{\partial z^+} + \frac{R}{2} \frac{\partial^2}{\partial x^2} \right) \psi(z^+,x) = 0.
\end{equation}

This equation represents an effective quantum-mechanical model for the wavefunction of transverse spacetime position. It is just a simple Schr\"odinger wave equation, and has wavelike solutions.  Notice that even though it is a quantum mechanical equation, and $\hbar$ has not been set to unity,   this quantum theory does not depend on  Planck's constant $\hbar$; instead, it  depends only on $R$.

This simple result shows the close affinity of the projective quality of M theory--- the way the extra or virtual dimension in the null direction is encoded in spatial gradients in the wavefront--- with the simple arguments from wave theory used in \cite{Hogan:2007pk,Hogan:2008zw} to
derive the properties of holographic indeterminacy. The transverse correlations of M theory are the same as those of a quasi-monochromatic ``carrier'' field of wavelength $R$. The basic properties can be derived from diffraction theory in a general way and seem likely to apply to any holographic theory.  The well-known close relationship of the Schr\"odinger  equation with the diffusion equation also makes it easy to see why the decoherence or spreading of a position wavepacket has the character of a transverse random walk--- about an $R$-length per $z^+$-time in each transverse dimension--- although the more general and accurate description is the wavefunction solution given here.

The quantum-mechanical wavefunction solution of Eq. (\ref{evolution}) along the longitudinal direction has  eigenfunctions of the form 
\begin{equation}
\psi_\omega(x)= \exp[ -i \omega z^+ + ik x)],
\end{equation}
with a dispersion relation 
 \begin{equation}
k_\omega=\pm \sqrt{2\omega/R}.
\end{equation}

Because it is a lightsheet theory, the  low frequency 
 modes describe  large longitudinal $(z)$ spatial separations in the emergent space. Position wavefunctions at longitudinal spatial wavelengths $L$ are described by eigenmodes with frequency $\omega= 1/L$ and have a ``fuzziness'' in the $x$ direction. 
The general solution for the transverse $x$ wavefunction can be written as a  sum,
\begin{equation}\label{wavefunction}
\psi(z^+,x)= \sum_L a_{L\pm} \exp[-iz^+/L \pm ix\sqrt{2/LR}].
\end{equation}
This behavior imposes a particular new kind of nonlocality on the emergent spacetime--- a coupling of the longitudinal and transverse positions even at large separation. These modes differ from  solutions to the  3+1D wave equation in field theory; there,  the degrees of freedom are spatial plane wave modes that evolve  independently aside from local couplings.  Note that this nonlocality is \emph{not} strictly proportional to the 11d Planck scale $M_{\rm pl}^{-1}$, but rather to the radius of the M-dimension $R$; these two scales differ by a factor of the string coupling $g_s^{2/3}$.

\subsection{Quantum Indeterminacy  of Transverse Position}

In our 3+1D world, the macroscopically 2+1D M-sheet (which of course, in Matrix theory has many more internal degrees of freedom including other microscopic virtual dimensions) appears to be a wavefront moving at the speed of light. The width of the sheet, $R$, corresponds to the minimum interval of time in the 3+1D world. The width of the sheet does not Lorentz contract in different frames, but is observer independent.  The incompressibility noted above means that the internal degrees of freedom described by the matrices,  KK-modes in the M-direction,  while they can take many forms, cannot exceed one degree of freedom per Planck area.   The theory we have just discussed relates transverse positions in the sheet at different times or different longitudinal locations.  This picture shows why the simple wave theory of refs. \cite{Hogan:2007pk,Hogan:2008zw}, based on transverse correlations in the wavefronts of monochromatic radation of wavelength $\lambda=R$, provides a faithful description of the same macroscopic correlations, and    the same  holographic noise.  The ``carrier wave'' in that theory represents the motion of an M-sheet wavefront through space, with a wavelength identified with $R$.
 
  The wavefunction does not depend on the mass of a particle or body, but describes the position of any body within the  spacetime. Other quantum degrees of freedom corresponding to particle motion are encoded in   terms we have chosen to ignore. In other words this   only describes the new quantum mechanics of   the holographic geometry associated with  a macroscopic system, including its emergent spacetime.  It does not yield the same macroscopic  limit as field-theory quantization of the relativistic wave equation in a classical spacetime: here,   longitudinal and transverse dimensions of the wavefunction are not independent, as they are for  plane wave modes of field theory.  For this reason holographic theories display a new and distinctive phenomenology.  Transverse positions have an indeterminacy that grows with scale, a behavior not present in  classical limits taken using field theory. Moreover the degrees of freedom described are not  physical degrees of freedom of classical gravity, so they do not appear in a direct quantization of general relativity. They are physical degrees of freedom of physical systems since bodies can be in different positions in space; but the classical spacetime is invariant under these shear perturbations, at least to first order.  Indeed this model is built on a flat spacetime background which remains unperturbed; the new effect is only  in the relative transverse positions of material bodies, built out of the  D0-branes represented by the matrices. At the same time, violations of Lorentz invariance, such as energy-dependent propagation velocity, vanish by construction: massless particles in holographic geometry always travel at exactly the speed of light, so there are no longitudinal dispersive spacetime propagation effects, even for very distant sources.

\subsection{Holographic Noise in an Interferometer}
Holographic noise describes the indeterminacy inherent in a spacetime constructed out of such modes. It
  can be characterised by a    power spectral density for dimensionless shear \cite{Hogan:2007pk,Hogan:2008zw},
\begin{equation}\label{psd}
S_H=R/2,
\end{equation}
independent of frequency.
This spectrum is a universal property of holographic quantum geometry, independent of any apparatus used to measure it. The following remarks concern  the response of an apparatus to the noise.

It is useful to consider a specific  measurement apparatus, and the quantum description of the system including its enclosed macroscopic spacetime:   a simple Michelson interferometer in flat spacetime.
The instrument is best described in a basis state where $z^+$ is aligned with one of the arms and $x$ with the other.   
 The  observable quantity is the laser intensity at the dark port.  This observable  corresponds to states oriented geometrically with the  incoming laser wavefront plane, so the signal phase is described by the transverse position  wavefunction, Eq. (\ref{wavefunction}). More specifically, we interpret $x$ to be the quantum operator for position of the beamsplitter as measured by the signal at the dark port.  The longitudinal direction is that of the incoming laser.    Wavefront phase in this direction serves as the reference clock in the Hamiltonian evolution.   The interferometer signal phase records the difference of position, transverse to this  incoming  null wavefront direction, of a single body (the beamsplitter) relative to itself  at a different time or longitudinal position--- corresponding to times separated by integer multiples of twice the arm length, $T=2L_0$.

The apparatus  shapes the spectrum of  the coefficients  $a_\omega$ (or equivalently, $a_L$).
The dark port field amplitude depends on the sum of  the field  for different null paths.
 In distance units the phase difference is the difference in  the transverse position of the beamsplitter between two times, separated by time $T$ as measured in the incoming-wavefront frame. The  position state at low frequencies is dominated by modes with  frequencies close to $\omega=1/T$, or $L=2L_0$.
There has to be significant power at $\omega=1/T$ in order for the spacetime state to extend to the apparatus size and encompass the physical separation of the optical elements. At the same time
the signal depends not at all on the position of system relative to the rest of the universe outside the apparatus.  The size of the cavity thus imposes a high pass filter on the modes describing the observable geometry of the system: $a_\omega=0$ for $\omega<1/T$.  

The signal phase difference between two times separated by $\Delta t$  wanders with time  for $\Delta t<T$: the mean transverse displacement grows with time.  This can be described as a spreading of a the overall wavepacket due to the random mixing of phases of the   modes of different $\omega$ in the sum, Eq.(\ref{wavefunction}).    
The position variance at large time intervals is insensitive to the high frequency spectrum and depends only on the lowest frequency present.  Assume that there is only one frequency, and  the apparatus is in a stationary state with $\omega=1/T=1/2L_0$:
\begin{equation}
\psi=e^{-it/T}\exp[i x/\sqrt{RL_0}].
\end{equation}
 In this state, the probability distribution of $x$ is flat: it is equally likely to take any value in the interval
\begin{equation}\label{interval}
-\pi \sqrt{RL_0}< x <\pi \sqrt{RL_0}.
\end{equation}
At large time separation $\Delta t \gg T$,   samples of position difference at $\Delta t>T$ are random variables drawn from this fixed probability distribution. Even for a more general spectrum with high pass cutoff, the position measured at times separated by more than $T$ is an independent random variable drawn from a distribution with the variance of a uniform distribution of width (\ref{interval}). Because there is actually a mix of  high frequencies, describing intermediate scales of the apparatus,   wavepacket spreading on short timescales resembles a random walk of beamsplitter position.   The lack of low frequencies thus results in an apparent ``bounded random walk'' of detectable position variation.

This effective theory fixes the predicted holographic noise in interferometer signals.   The prediction has no parameters aside from the value of $R$.
Put another way, measurements of holographic noise directly measure $R$, independent of the Planck length as measured from gravity.   

 The effective wave theory can be adapted to predict the noise in more general interferometer setups. The   interferometer currently most sensitive to holographic noise is the interferometric gravitational wave detector  GEO600, which indeed displays unexplained noise in its most sensitive band ($\approx$ 500 to 1300 Hz)  that approximately agrees with\footnote{In ref. \cite{Hogan:2008zw} it was estimated incorrectly that the apparent spectrum in this device should show an upturn below the frequency ($\approx$ 550 Hz)  corresponding to the effective averaging time in the power-recycled cavity; in fact the calibrated spectrum should just continue to be flat with the value in Eq. (\ref{psd}). We are grateful to the GEO600 team for clarifying this point.} the zero-parameter prediction, Eq. (\ref{psd}). In LIGO, unlike GEO600,  the interferometer arms have separate Fabry-Perot cavities that do not include the beamsplitter.  A gravitational wave strain has an effect on the signal that increases linearly with the finesse of these cavities, increasing the effective length of the arms.  We  guess that  the  rms  holographic displacement of the beamsplitter increases only as the square root of the effective arm length, so instead of measuring  the universal spectrum (Eq. \ref{psd}), LIGO measures an ``apparent'' holographic noise spectrum $S_H^{1/2}$  reduced by the square root of the cavity finesse. This is  below the current sensitivity level but should be detectable with future improvements.  (If holographic noise is a real effect, future gravitational wave detectors will require redesign to mitigate the sensitivity  limitation from holographic noise.)

\section{Matrix Scattering Theory}
\begin{figure}
\begin{center}
\includegraphics[width=2.5in]{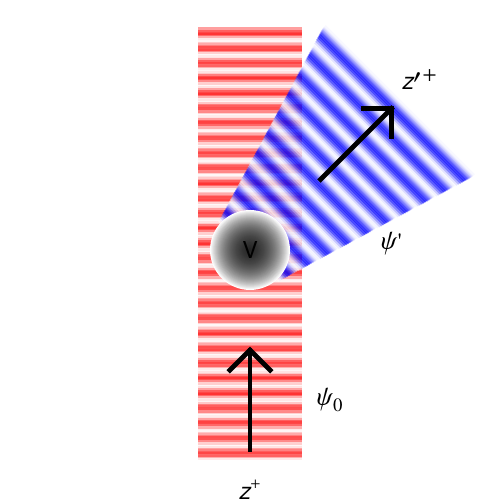}
\caption{Scattering in Matrix Theory. The sketch shows a system in a lab frame; the wavefronts are separated by approximately $R$.  When the original wavefunction $\psi_0$ is scattered by a potential $V$, the resultant wavefunction $\psi'$ will generally propagate in a different lightcone direction from the original, and possesses an increasingly uncertain transverse position.} \ \\\end{center}
\end{figure}
The effective theory  above provides a bridge  connecting Matrix Theory and phenomenology, by identifying a new kind of normal quantum mode for macroscopic spacetime. However, it does not describe all the physics contained in the Matrix theory. In what follows we pursue the Matrix description  in more detail. Analysis of this kind may be motivated in some conceivable experimental situations, for example in describing holographic uncertainty in the relative positions of bodies lighter than a Planck mass.

The novel behavior of Matrix theory becomes evident upon examining the scattering of a holographic wavefunction.  We emphasize that a wavefunction of the theory is \emph{not} a wavefunction of a field which depends on spacetime coordinates, \emph{it is a wavefunction of the spacetime coordinates $X^i$ themselves}.  This is  a quantum description, arising in Matrix Theory, of a macroscopic system, including its emergent spacetime.  As noted in the context of the effective theory above, it does not yield the same classical limit as field-theory quantization in a classical spacetime nor direct quantization of General Relativity.

 Just as in standard quantum mechanics whereby the measurement of an observable will change the wavefunction's basis, interaction with a potential will change the choice of longitudinal versus transverse coordinates.  Unfortunately, scattering theory in the lightcone is not as straightforward as in standard quantum mechanics, since it is precisely this change of the propagation (`time') coordinate which makes a perturbative analysis awkward \cite{Polchinski:1997pz}.  Nonetheless, one can anticipate an approximate solution.  Consider again the lightcone Hamiltonian corresponding to the Lagrangian (\ref{lagrangian}), which we now more appropriately label as the generator of translation in the $z^+$-direction,
\begin{eqnarray}
\nonumber
{\hat \Pi}_{+} &\equiv& E-p^z \\
\label{hamiltonian}
&=& R {\rm tr} \ \sum_i \left\{ \frac{ {\hat \Pi}^2_i }{ 2} + {M_{\rm pl}^6 \over 4} \sum_j [{\hat X}^i,{\hat X}^j]^2 \right\}
\end{eqnarray}
 as well as that of the generator of $z^-$-translation,
 \begin{eqnarray}
\nonumber
{\hat \Pi}_{-} &\equiv& E+p^z \\
&=& {\hat \Pi}_{+} + \frac{2N}{R}.
\end{eqnarray}
We will be considering wavefunctions $\psi$ which are eigenstates of ${\hat \Pi}_{+}$, which are then automatically eigenstates of ${\hat \Pi}_{-}$, and thereby have the form
\begin{equation}
\label{psi0}
\psi( X^i, z^+,z^-) = e^{-ip_+z^+ - i (p_+ + 2N/R) z^-} \psi(X^i,0,0). 
\end{equation}
For this to be a reliable calculation requires ``small lightcone energy," $p_+ \ll N/R$.  We would expect that any scattered wavefunction $\psi'$ obeys the usual Hamiltonian equations of motion \emph{in the new direction} of propagation $z^{'+}$ generated by  ${\hat \Pi}'_{+}$ which is now a function of new transverse coordinates $X^{'i}$, up to a source term $J$,
\begin{equation}
\label{eom}
\left[ i \frac{\partial}{\partial z^{'+}} -  R {\rm tr} \ \sum_i \left\{- \frac{ 1}{ 2} \frac{\partial^2}{\partial X^{'i2}} + {M_{\rm pl}^6  \over 4} \sum_j [X^{'i},X^{'j}]^2 \right\} \right] \psi'(X^{'i},z^{'+},z^{'-}) = J(X^{'i},z^{'+},z^{'-}).  
\end{equation}
This is schematically illustrated in Figure 1.  The source is given by the original wavefunction $\psi_0$ in the presence of some potential $V$, normalized with the same factor of $R$, but cast into the new coordinate system by appropriate rotation and tracing.  The new coordinates $(X^{'i}, z^{'+},z^{'-})$ can of course be obtained by some linear transformation upon two spatial coordinates, using constants $a,b,c,d$ such that $ad-bc = 1$:
\begin{eqnarray*}
x'  &=& a x + b z, \\
z'  &=& c x + d z, \\
t' &=& t.
\end{eqnarray*}
To effect this transformation in lightcone coordinates, however, is more subtle and involves expressing any $z$-dependence in terms of the lightcone coordinates $z^{\pm}$, and similarly for $z^{'\pm}$:
\begin{eqnarray*}
x' &=& a x + b z^+ - bz^-, \\
z^{'+} &=& c x +  (d+1)z^+ - (d-1) z^-, \\ 
z^{'-} &=&- c x- (d-1) z^+ + (d+1) z^-. \\ 
\end{eqnarray*}
The source can now be written as
\begin{eqnarray*}
 J(X^{'i},z^{'+},z^{'-}) &=& R \int dX^i dz^+ dz^- \  \delta \left( X^{'i} - a X^i - bz^+ I +b z^-  I \right) \delta \left( z^{'+} -  cx -(d+1) z^+ + (d-1)z^- \right)  \\
&& \cdot \delta \left( z^{'-} + cx + (d-1)z^+ - (d+1) z^- \right)  V(X^i,z^+, z^-) \psi_0(X^i,z^+,z^-) .
\end{eqnarray*}
The spacetime dependence of the potential $V$ deserves comment.  While $V$ can have arbitrary dependence on each D0-brane's transverse position $X^i$, its dependence upon the propagation direction can only be on the center of mass coordinates $z^{+},z^{-}$, requiring that we replace $Z \rightarrow zI, T \rightarrow tI$ and then switch to lightcone coordinates.  In the special case that there is no dependence upon the propagation direction $z^+$ this reduces to the familiar time-dependent Schr\"odinger's equation in quantum mechanics.  In any case, (\ref{eom}) is easily inverted to yield
\begin{equation}
\label{psiprime}
\psi'(X^{'i},z^{'+},z^{'-})  =  \int d{\tilde X} ^{'i} d {\tilde z}^{'+} d{\tilde z}^{'-} \ G \left( X^{'i},z^{'+},z^{'-}; {\tilde X}^{'i},{\tilde z}^{'+},{\tilde z}^{'-} \right) J({\tilde X}^{'i},{\tilde z}^{'+} ,{\tilde z}^{'-})
 \end{equation}
where the Green's function is defined to include enforcement of the $p_- = p_+ + 2N/R$ constraint:
\begin{eqnarray*}
G \left( \Delta X^{i}, \Delta z^{+}, \Delta z^{-} \right) &=& \int \frac{ dp_+ dp_- dp_i }{(2 \pi)^{11}} e^{-ip_+ \Delta z^+ - i p_- \Delta z^- + i  {\rm tr} \ p_i \Delta X^i} {\tilde G}(p_+, p_i) \delta \left( p_- - p_+ - \frac{2N}{R} \right) \\
&=& \int \frac{ dp_+  dp_i }{(2 \pi)^{11}} e^{-ip_+ \Delta z^+ - i (p_+ +  \frac{2N}{R}) \Delta z^- + i  {\rm tr} \ p_i \Delta X^i} {\tilde G}(p_+, p_i).
\end{eqnarray*}
This means that when acted upon by the Hamiltonian in position-space, the Green's function yields the expected $\delta$-function up to a phase, 
\[ {\hat \Pi}_+(X^{i},z^{+}) G \left( \Delta X^{i}, \Delta z^{+}, \Delta z^{-} \right)  = e^{-2 i N \Delta z^{-}/R} \delta \left( \Delta z^+ + \Delta z^- \right) \delta \left( \Delta X^i \right) . \]
Since we won't know what the new direction of propagation is until \emph{after} the scattering process, the expression (\ref{psiprime}) may appear to have limited utility.  However, let us continue, and apply this formalism to the example of interest.
\section{Beamsplitters and Interferometers in Matrix Theory}
\begin{figure}
\begin{center}
\includegraphics[width=2.5in]{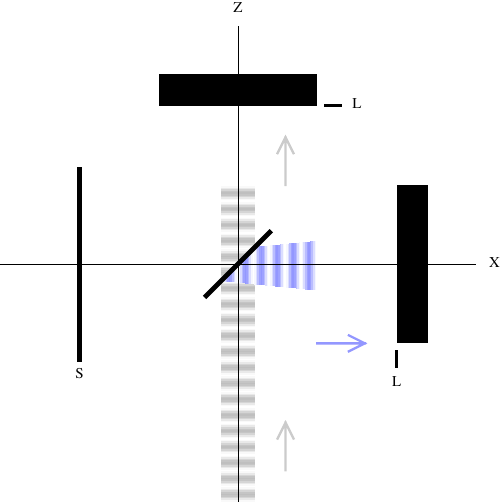}
\hspace{0.5in}
\includegraphics[width=2.5in]{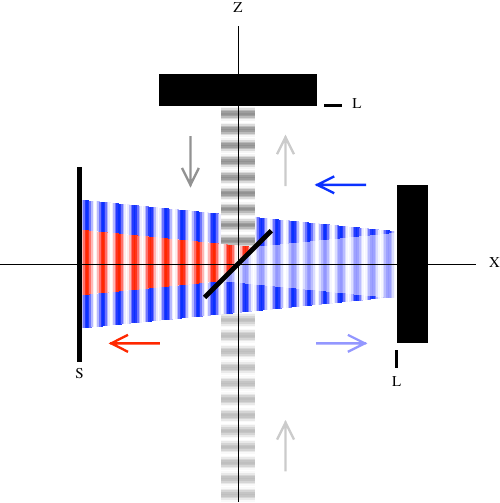}
\caption{Interferometer setup.  A wavefunction enters from below, and can reach the surface $S$ via two paths which differ in their order of beamsplitter interaction.  The wavefunction which scatters first will have much more time to acquire a large uncertainty in the transverse position.  The wavefunction which scatters second has much less transverse uncertainty.} \ \\
\end{center}
\end{figure}
To demonstrate the existence of holographic noise, we follow the path of such a Matrix Theory wavefunction in a transverse-measurement beamsplitter, as shown in Figure 2.  The wavepacket (\ref{psi0}) arrives from $z^+ = -\infty$, reflects off of a mirror at $z=L$, and then returns.  This mirror is represented by the potential
\begin{equation}
V_{z-mirror} = \left\{
\begin{array}{rr}
0, & \hspace{0.5in} Z < L, \\
\infty, & \hspace{0.5in} Z \ge L.
\end{array}
\right.
\end{equation}
This is of course not a small perturbation but rather imposes the boundary condition $\psi_0(z = L) = 0$, meaning that the reflected  wavefunction is identical but propagating in the $z^-$-direction.  The combined (forward and reflected) initial wavefunction is then determined to be
\begin{equation}
\label{psi}
 \psi_0 (X,Y,z^+,z^-) = \left[ e^{-ip_+(z^+ -L) - i (p_+ + 2N/R) (z^- + L)} - e^{-ip_+ (z^-+L) - i (p_+ + 2N/R) (z^+-L)}  \right] \psi_0(X,Y,0,0). 
 \end{equation}
We will first focus on the forward component of this, since this is the part that will scatter first.

So as to maximize the effect of induced uncertain transverse position we assume the scattered wavepacket travels in the $+x$-direction, making the coordinate change simply
\[  z' = x,  \hspace{0.5in}  x' = - z. \]
We represent the beamsplitter by a potential which is both infinitely repulsive and infinitely thin located along the line given by $x=z$, parameterized by some coupling constant $g$,
\[ V(X,Y,z,t) = g \delta(X-zI) . \]
This produces a source term which is (as expected) simply the rotation into the new coordinates:
\begin{equation}
\label{bssource}
J(X',Y,z^{'+},z^{'-}) = g R \delta \left( X' + z^{'+} I - z^{'-} I \right) e^{-ip_+(z^{'+} -L) - i (p_+ + 2N/R) (z^{'-}+L)}  \psi_0(X',Y,0,0).
\end{equation}
Plugging this into (\ref{psiprime}), the scattered wavefunction is
\begin{eqnarray}
\nonumber
\psi'(X',Y,z^{'+},z^{'-}) &=&  \int \frac{ d{\tilde p}_+ d{\tilde p}_- d{\tilde p}_X d {\tilde p_Y} }{(2 \pi)^{4}} e^{-ip_+ (z^+ - {\tilde z}^+) - i (p_+ +  2N/R) (z^- - {\tilde z}^- ) + i  {\rm tr} \ [ p_X (X-{\tilde X}) + p_Y (Y-{\tilde Y}) ]} \\
\label{scatteredwf}
 && \hspace{1in} \cdot {\tilde G}(p_+, p_X,p_Y) J({\tilde X}',{\tilde Y},{\tilde z}^{'+},{\tilde z}^{'-}) .
\end{eqnarray}
We see that measuring the wavefunction in this transverse basis has thrown away any information about $X'$ except for the center-of-mass, which is now identified to be $-z'$.  

We believe that the wavefunction now undergoes a diffusion-like process for the following reason.   Consider the term in ${\hat \Pi}'_+$ corresponding to the non-commutation of coordinates, $\sim [X',Y]^2$.  Since (\ref{bssource}) enforces that $X'$ is now proportional to the identity matrix, this term vanishes as $I$ commutes with everything:
\[ \langle \psi' | [ X', Y]^2 | \psi' \rangle = 0. \]
If we then neglect the corresponding term in the Green's function (\ref{scatteredwf}), we may estimate it as simply
\[ G(z^+, X, Y) \sim \left[ i\partial_+ + \frac{R}{2} \left( \partial_X^2 + \partial_Y^2 \right) \right]^{-1} . \]
This is the (inverse of the) diffusion equation, making contact with the previous section on the effective wave theory, and implying that the reflected wavefunction will spread as
\[ (\Delta X')^2 \sim L R . \]
The wavefunction will then diffuse as it travels the length $L$ until it hits the mirror, then travel another length $L$ until it hits the screen $S$.

Of course, spreading of the wavefunction is just indicative that it is not a single eigenstate of the Hamiltonian.  This happens in usual quantum-mechanical scattering, and will happen in our system even in the forward direction due to the crudeness of the perturbative method, whereas ideally a more sophisticated treatment of the scattering would remove this effect.  However, this effect should be suppressed at high energies for the same reason a slowly-varying potential will not induce transitions: in this `adiabatic' limit in which the wavelength $1/p_+$ is much smaller than the $z^+$-variation of the potential, any wavefunction spreading should be due entirely to the holographic change-of-basis which we have focused on in the present article.

We must now include the presence of the reflected portion of the wavefunction (\ref{psi}).  This results from the wavepacket passing through the beamsplitter during its forward path, then hitting the mirror located at $z = L$.  It will then reflect off this, traveling in the $-z$-direction until it encounters the beamsplitter again, where we assume it now scatters (if it doesn't, it will never hit the surface and thus is of no interest to us).  The scattering process is similar to (\ref{scatteredwf}), but the wavefunction will diffuse over the small distance to the detection surface $S$.  Thus, when these two wavefunctions (which began as identical, and are now different only in their order of scattering) are recombined on $S$ we will find one is much more spread out than the other.  This phenomenology resembles that derived  above from effective wave theory.

\section{Conclusions}

The holographic noise derived here has previously been shown to approximately agree with the spectrum of currently unexplained continuum noise in the best operating interferometer, GEO600, in its most sensitive band, above about 500 Hz. Therefore, either the effect has already been detected,  or it is within reach to rule it out using existing technology. In the latter case, it  should be possible to exclude a broad class of theories that encompass the idea of holography; the position noise is directly related to the holographic information bounds around which these theories are constructed. The properties of holographic noise, such as its exact spectrum and its  shear character, are more specific than more generalized phenomenological descriptions of quantum spacetime noise (e.g., \cite{AmelinoCamelia:2001dy,Smolin:2006pa}).

If the effect exists,  a diverse and detailed experimental program is motivated.  The effective 
theory for the degrees of freedom fixes both the  spectrum of  noise and the character of its geometrical correlations.  The spectrum of the noise depends only on $R$ and therefore provides a direct experimental window on the Planck scale. The expected signal correlations between nearby, but not coincident interferometers will be important to predict particular signatures that can distinguish holographic noise from other more prosaic sources: in-common signals occur, due to quantum wavefunction collapse, even between machines with no physical connection, and even on timescales too short to mimic other kinds of in-common disturbances, such as transmitted acoustic vibrations.  

Eventually, other technologies such as atom interferometers will approach the sensitivity required to study holographic noise at smaller scales.   Interactions in that case occur with bodies with individual masses less than a Planck mass, in which case  a correct calculation of the collective behavior of the wavefunction may require a fuller account of the matrix degrees of freedom.
 
\section{Acknowledgments}
 We would like to thank D. Kabat for extensive assistance regarding the Matrix Theory aspects of this work, and members of the GEO600 team (H. Grote, S. Hild, and H. L\"uck) for consultation about their detector.  Our work was supported by the Department of Energy.

\end{document}